\def \pom {{\scriptscriptstyle \kern -0.1em I \kern -0.25em P}}
\input epsf.tex
\def\DESepsf(#1 width #2){\epsfxsize=#2 \epsfbox{#1}}
\documentstyle{aipproc}
\begin{document}
\title{Diffraction in DIS and Elsewhere}

\author{Davison E.~Soper
\thanks{Research supported by the United States Department of
Energy.}}
\address{Institute of Theoretical Science, University of Oregon
Eugene, Oregon 97403}

\maketitle

\bigskip

\leftline{Talk at DIS97 Conference, Chicago, April 1997}

\begin{abstract}
I review some of the results presented in the working group on
diffraction at DIS97, with a particular emphasis on the theory of
diffractive hard scattering.
\end{abstract}

\section*{Introduction}

This is one of two talks summarizing the results of the working
group on diffraction at DIS97. The other talk, by Paul Newman
\cite{Newman}, should be read in conjunction with this one. My aim is
mainly to provide some theoretical background for interpreting the
experimental results and to point out some connections among the
results. The talk by Newman summarizes the experimental results from
$e^+p$ scattering. This talk describes some highlights from the
results from $\bar p p$ scattering.  

\section*{Theory of Diffractive DIS}

I begin with some theoretical background on diffractive deeply
inelastic scattering (DIS). Consider the process $e + p \to e + p +
X$. Denote the initial lepton momentum by $\ell^\mu$, the final
lepton momentum by $\ell^{\prime\,\mu}$, the initial proton momentum
by $p^\mu$, the final proton momentum by $p^{\prime\,\mu}$, and the
momentum of the photon or $Z$ boson emitted by the lepton by $q^\mu =
\ell^\mu - \ell^{\prime\,\mu}$.

It will be helpful to have in mind a definite reference frame, or
rather a definite set of reference frames. Let us denote components
of four-vectors by $v^\mu = (v^+,v^-,{\bf v}_T)$, where $v^\pm = (v^0
\pm v^3)/\sqrt 2$. Then we choose the frame such that ${\bf q}_T =
{\bf p}_T =0$. In such a frame, the components of $p^\mu$ and $q^\mu$
are $p^\mu = (p^+, m^2/(2 p^+), {\bf 0})$ and $q^\mu \approx (- x p^+,
Q^2/(2 x p^+), {\bf 0})$. Here I use the usual DIS variables $Q^2 = -
q^2$ and $x = Q^2/(2\,q\cdot p)$. As usual in DIS, $Q$ should be
large compared to 1 GeV. It is often helpful in analyzing DIS to let
$p^+$ be large, on the order of $Q$. However, sometimes one gains
insight by taking $p^+$ to be finite, of order $m$. For example, we
can take $p^+ = m/\sqrt 2$, so that we are in the proton rest frame.
Some of the subsequent discussion will concern this frame switching. 

In the diffractive process considered, we observe the scattered
proton, with momentum components 
\begin{equation}
p^{\prime \mu} = ((1-x_\pom)p^+,
{m^2 + ({\bf p}^\prime_T)^2
\over 2 [1-x_\pom] p^+}, {\bf p}^\prime_T)
\end{equation}
The fraction of the proton's longitudinal momentum that is lost in
the scattering, $x_\pom$ should be fairly small, say  $x_\pom
\lesssim 0.05$. The kinematics requires $x \le x_\pom$. Thus $x$ must
also be small. The invariant momentum transfer,
\begin{equation}
t \equiv (p - p^\prime)^2 = - {{\bf p}_T^2 + x_\pom^2 m^2 \over
1-x_\pom}, 
\end{equation}
should be finite, $|t|\lesssim 1\ {\rm GeV}$.

We should note that in current experiments, one often omits the
detector for the scattered proton and substitutes a rapidity gap
signal. In this case, $t$ is not observed; the corresponding
theoretical formulas should then be integrated over $t$, with the
assumption that $|t|\lesssim 1\ {\rm GeV}$ dominates the integral.

The summary statement of these kinematics is that, from the point of
view of the proton, the scattering is rather gentle, although it is
probed by a very hard virtual vector boson. This gentle scattering
is often attributed to ``pomeron exchange.'' However, I will not
make use of the Regge theory that incorporates this language for
some time in this talk.

\begin{figure}[htb]
\centerline{\DESepsf(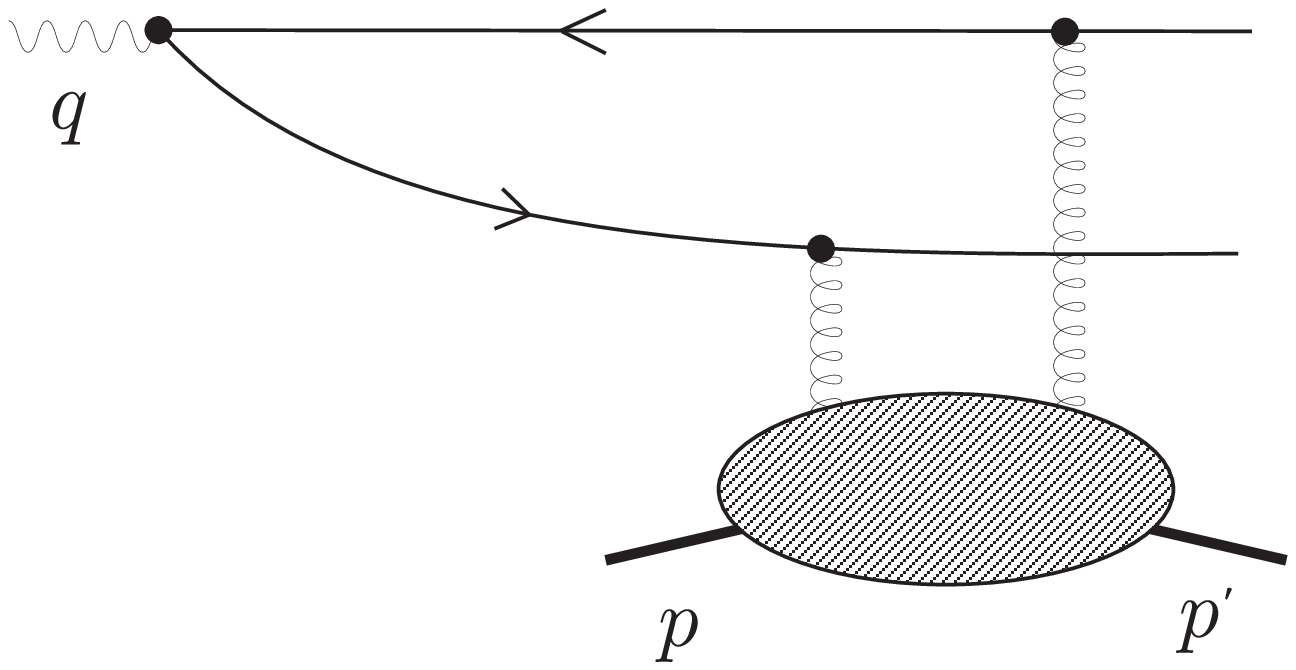 width 7 cm)
\DESepsf(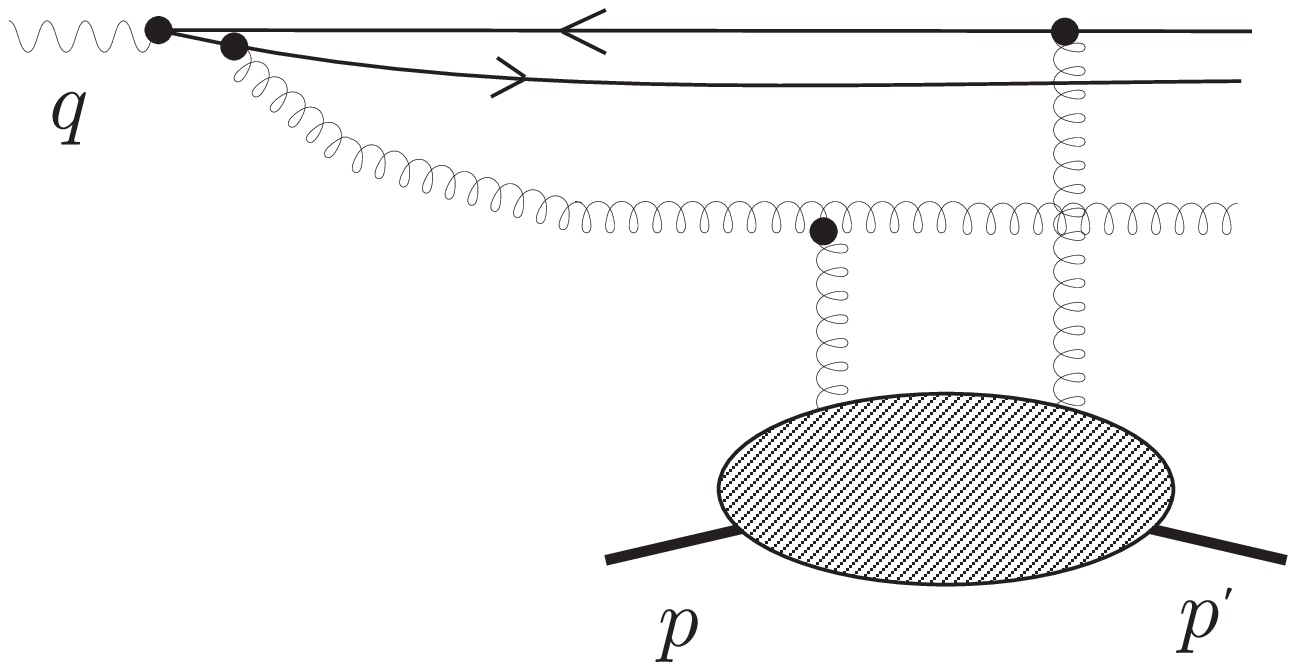 width 7 cm)}
\vspace{10pt}
\caption{Two graphs for a model of diffractive DIS.}
\label{F:Nikolaev}
\end{figure}

What is the microscopic origin of diffractive DIS? Let us look at it
in the proton rest frame. Then $q^-$ is very large, of order
$Q^2/(x m)$, while $q^+$ is of order $x m$. Thus one can think of the
vector boson as a system of quarks and gluons moving with very large
momentum in the minus direction. In the simplest model, this system
of quarks and gluons interacts with the proton via the exchange of
two soft gluons, as illustrated in Fig.~\ref{F:Nikolaev}. This model
has been extensively analyzed by N.~N.~Nikolaev and collaborators
and is discussed in his talk \cite{Nikolaev}. See also the talk of
W\"usthoff \cite{Wusthoff}.

\begin{figure}[htb]
\centerline{\DESepsf(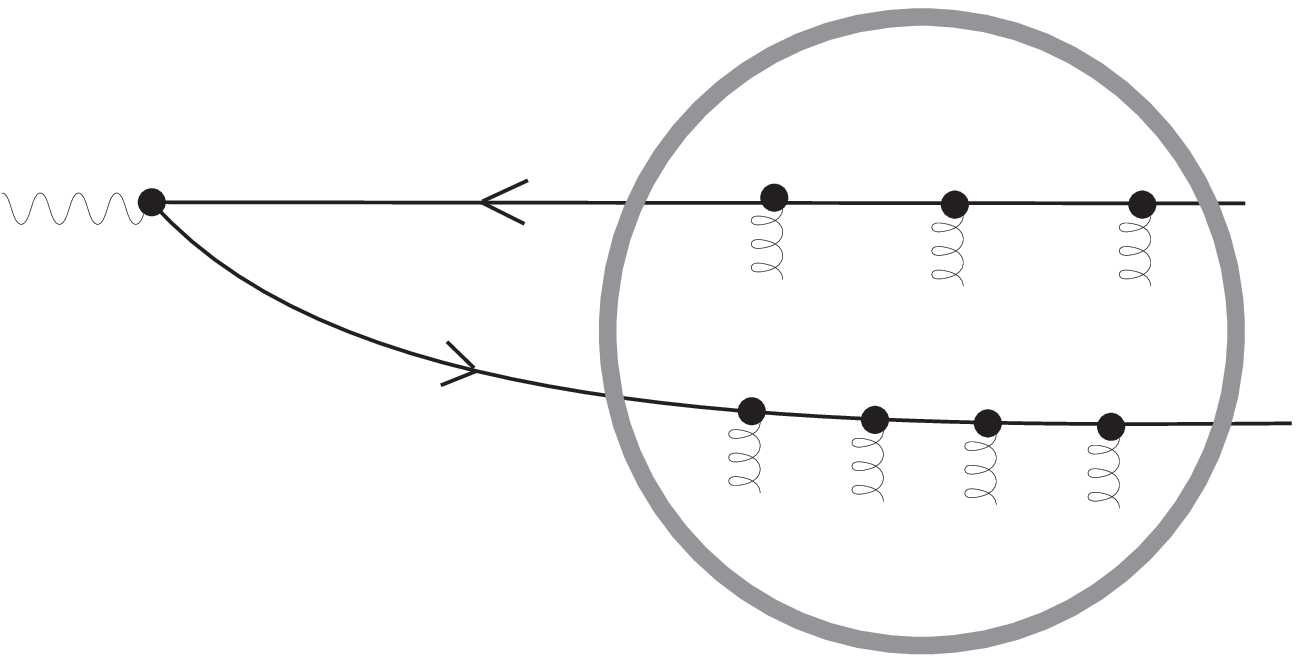 width 7 cm)
\DESepsf(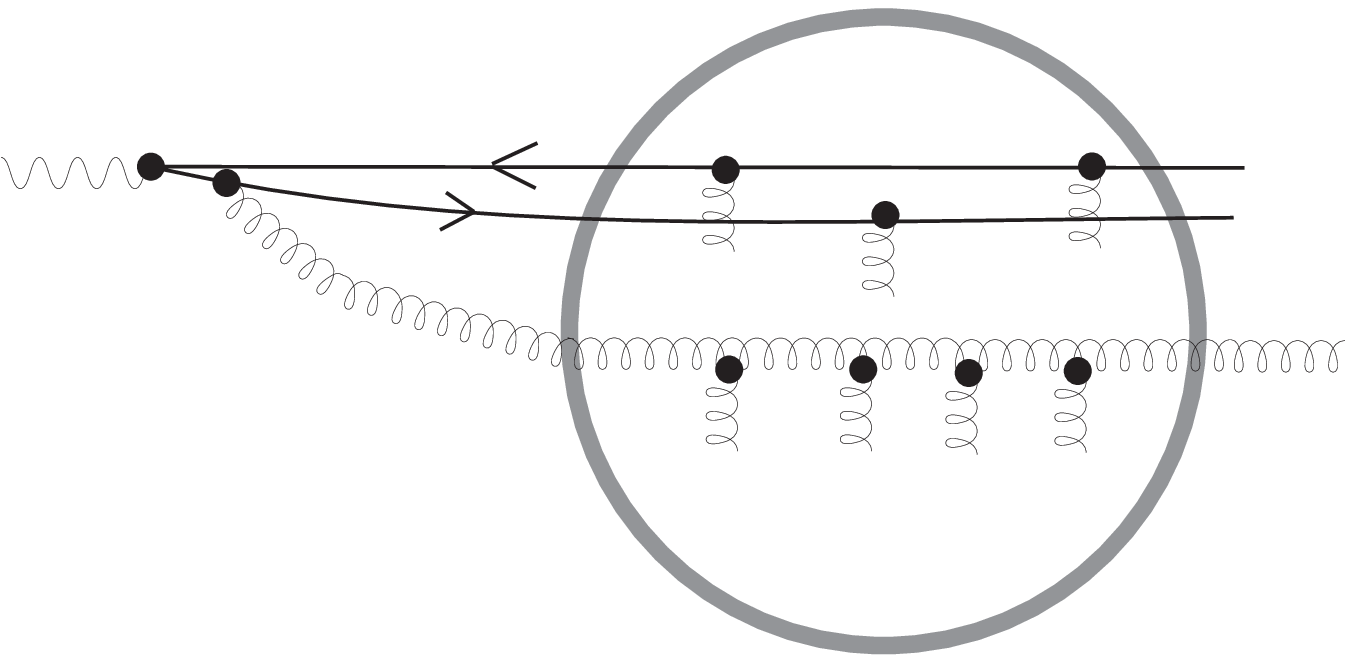 width 7 cm)}
\vspace{10pt}
\caption{Another picture of diffractive DIS.}
\label{F:Hebeker}
\end{figure}

Another version of this picture, illustrated in Fig.~\ref{F:Hebeker},
has been proposed by Buchm\"uller, Hebeker, and McDermott and is
discussed in the talk of Hebeker \cite{Hebeker}. The proton at rest
is represented by a circle. Partons in the virtual boson wave
function pass through the soft color field of the proton and undergo
a color rotation. If the net color transfer is color {\bf 1}, then
the proton has a chance to remain intact. The difference between
this and the model in  Fig.~\ref{F:Nikolaev} is that one expects
more than just two soft gluons to be exchanged.

\begin{figure}[htb]
\centerline{\DESepsf(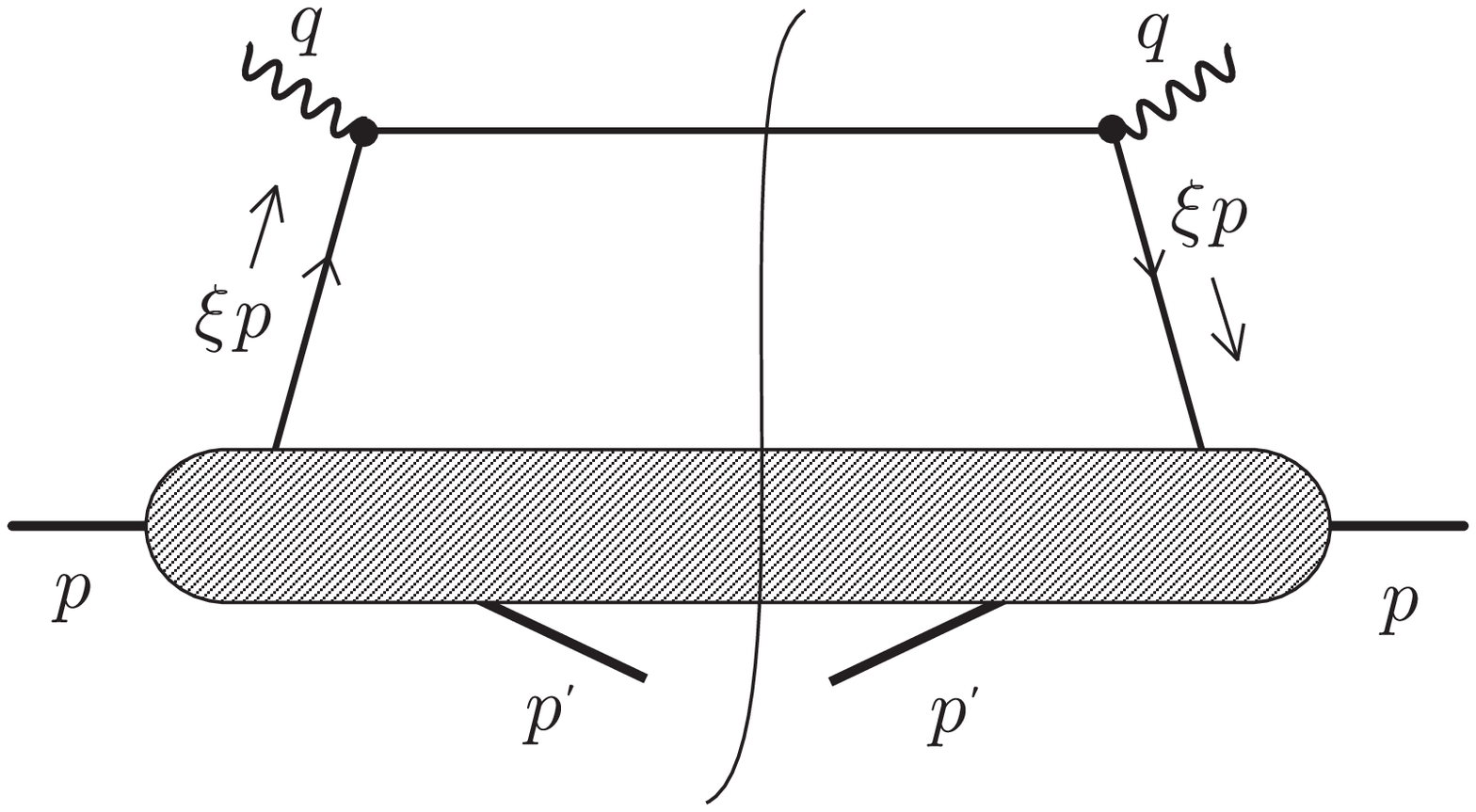 width 7 cm)
\DESepsf(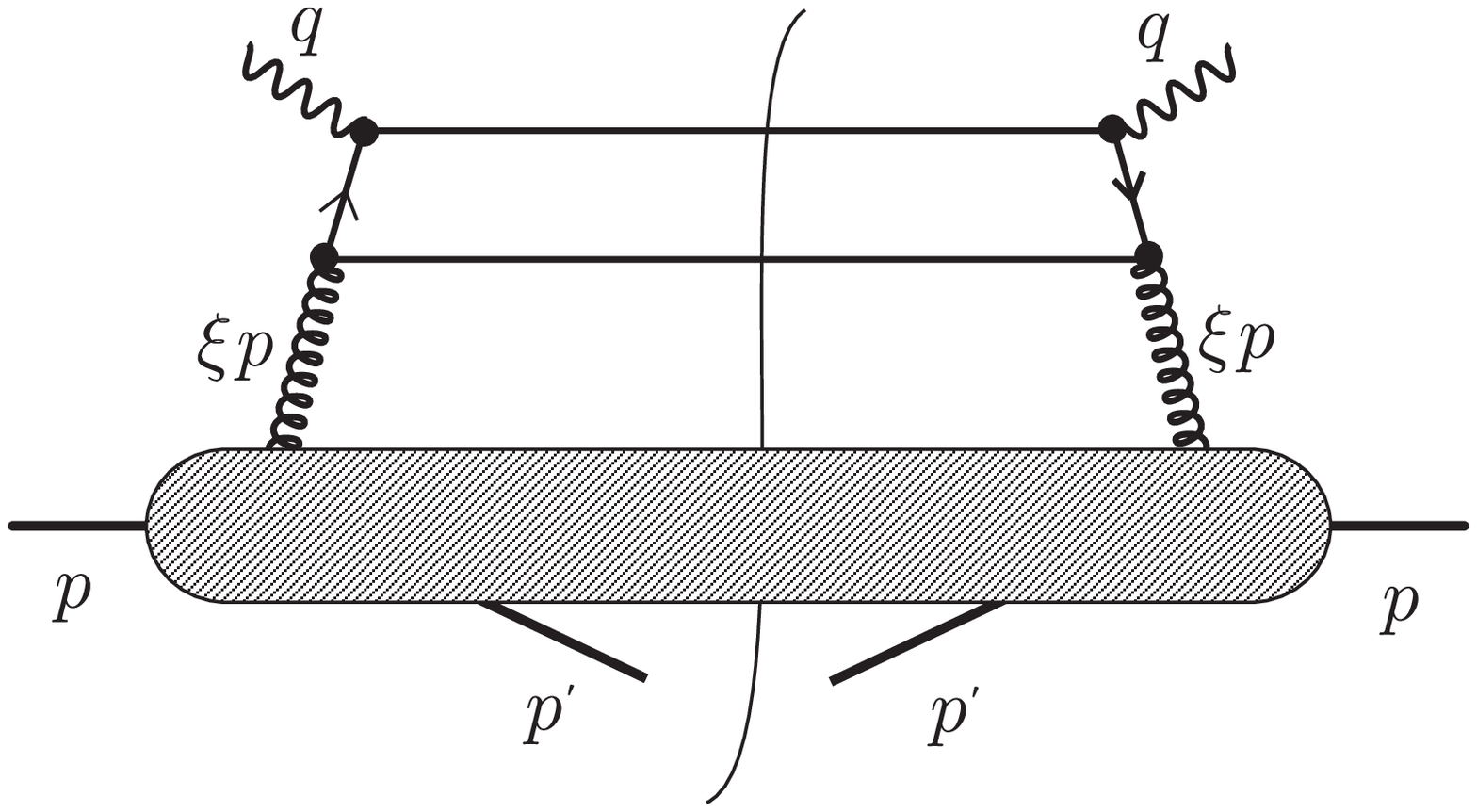 width 7 cm)}
\vspace{10pt}
\caption{Diffractive factorization in DIS.}
\label{F:diffractiveDIS}
\end{figure}

There is another view of diffractive DIS, as discussed in the talk of
A.~Berera \cite{Berera}. In this view, it is convenient to employ a
reference frame with $p^+$ of order $Q$. The fast moving proton is
viewed as being composed of partons. One of these partons, which
carries a fraction $\xi$ of the proton's plus momentum, is scattered
by the virtual photon. Although a parton has been removed from the
proton, there is some probability that a proton will reform from
the debris, having been scattered with a momentum transfer
$(x_\pom,t)$. The function that gives this probability is analogous to
the ordinary inclusive parton distribution function and is called the
diffractive parton distribution function, ${d f^{\rm
diff}_{a/p}(\xi;x_\pom,t;\mu) /(dx_\pom\,dt)}$. The measured
diffractive structure function ${d F_2^{\rm diff}(x,Q^2;x_\pom,t)/
(dx_\pom\,dt)}$ is then a convolution of the diffractive parton
distribution function with the {\it same} hard scattering function
$\hat F_{2,a}\left({x / \xi},Q^2;\mu\right)$ that is used to describe
ordinary inclusive DIS:
\begin{equation}
{d F_2^{\rm diff}(x,Q^2;x_\pom,t) \over dx_\pom\,dt}\sim
\sum_a\int_0^{x_\pom}\!\!d\xi\
{d f^{\rm diff}_{a/p}(\xi;x_\pom,t;\mu) \over dx_\pom\,dt}\
\hat F_{2,a}\left({x / \xi},Q^2;\mu\right).
\label{difffact}
\end{equation}
This equation expresses the hypothesis of {\it diffractive
factorization}. It is illustrated in Fig.~\ref{F:diffractiveDIS},
in which the left hand cut diagram shows the Born contribution to
$\hat F_2$ and the right hand diagram shows a next-to-leading-order
contribution, with a gluon as the parton entering the hard
interaction. 

Eq.~(\ref{difffact}) represents the model introduced by Ingelman and
Schlein \cite{IngelmanSchlein}, minus its Regge content. That is, the
diffractive factorization hypothesis does not say anything about a
relation of the diffractive parton distributions to Regge theory. If
we add Regge theory, then the form of these distributions is
restricted. 

Diffractive factorization was introduced in Refs.~\cite{BereraSoper}
and \cite{Veneziano32}. For lepton-hadron processes, diffractive
factorization appears to be a consequence of QCD \cite{BereraSoper}.
The general perturbative argument is supported by a one loop
calculation of Graudenz\cite{Graudenz}.

The diffractive parton distributions have a specific definition in
terms of matrix elements of certain operators \cite{BereraSoper}.
They obey the same DGLAP evolution equation as the ordinary inclusive
parton distribution functions. (This statement is self evident once
one accepts Eq.~(\ref{difffact}): the measured $F_2$ does not depend
on the factorization scale $\mu$, so the dependence of the parton
distribution on $\mu$ must be just that required to cancel the $\mu$
dependence of the calculated hard scattering function $\hat F_2$.)

The diffractive parton distributions are non-perturbative objects
that must be determined from experiment. Indeed, they have been
determined from the HERA experiments. One of the determinations, by
Alvero, Collins, Terron and Whitmore, was discussed in the talk of
Whitmore \cite{Whitmore}. This determination makes use of the form for
the functions suggested by Regge theory. The determinations are not
very precise, but there are some qualitative surprises. First, the
partons produced in diffractive scattering appear to be predominantly
gluons. Second, it appears that often a big fraction of the available
momentum is carried by a single gluon.

A caveat is in order. From the talks of Bartels \cite{Bartels} and of
Nikolaev \cite{Nikolaev} and from the discussions, we learned to
beware of the region $(1-\beta)Q^2 \lesssim M^2$, where $\beta =
x/x_\pom$ and $M$ is a mass scale of order 1 GeV. In this region,
power corrections, suppressed by a factor $M^2/([1-\beta]Q^2)$ may be
important.

Figs.~\ref{F:Hebeker} and \ref{F:diffractiveDIS} illustrate two
pictures of diffractive DIS that appear to be quite different.
We learned from the talk of Hebeker \cite{Hebeker} that, although
these pictures emphasize different features of the interactions, they
are physically the same. They are related by making a Lorentz
transformation from a frame in which the proton is at rest to a frame
in which it has a large momentum. One must then use gauge invariance
and make some appropriate approximations to see the correspondence.
The basic idea is illustrated in Fig.~\ref{F:correspondence} below.
In the left hand picture, the quark labeled $k$ travels forward in
time from the boson vertex to the interaction with a gluon from the
proton. In the right hand picture, because of the Lorentz
transformation, the (anti-) quark travels forward in time from the
interaction with a gluon to the interaction with the vector boson. In
the left hand picture, the quark appears to be a constituent of the
vector boson. In the right hand picture, its antiparticle appears to
be a constituent of the proton.

\begin{figure}[htb]
\centerline{\DESepsf(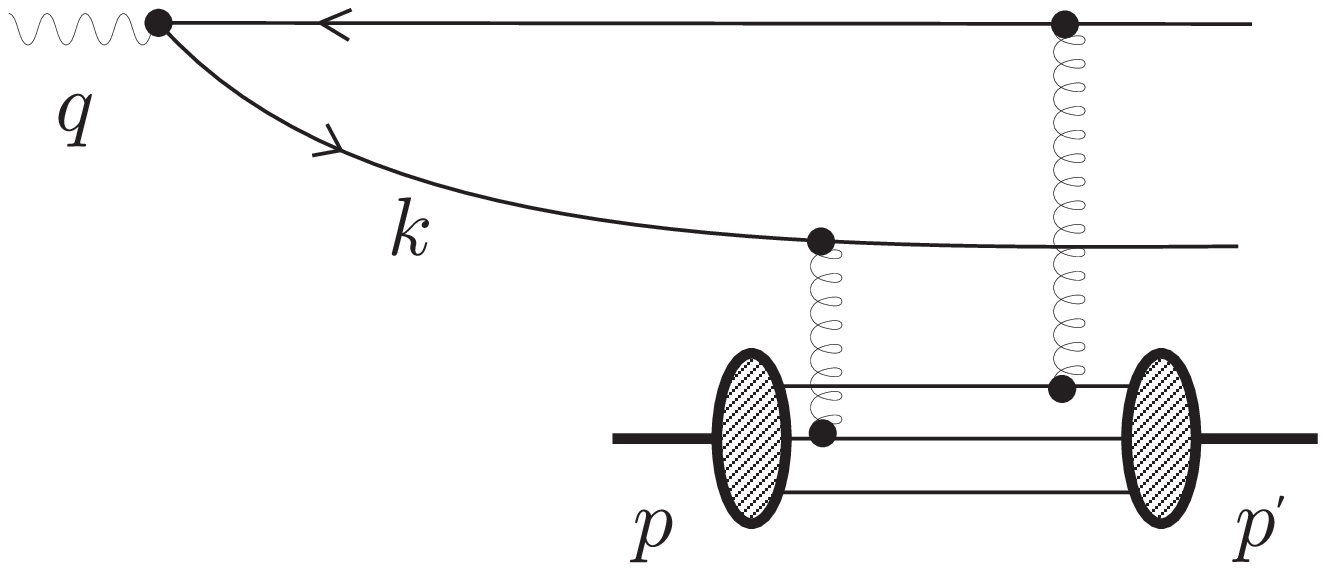 width 7 cm)
\DESepsf(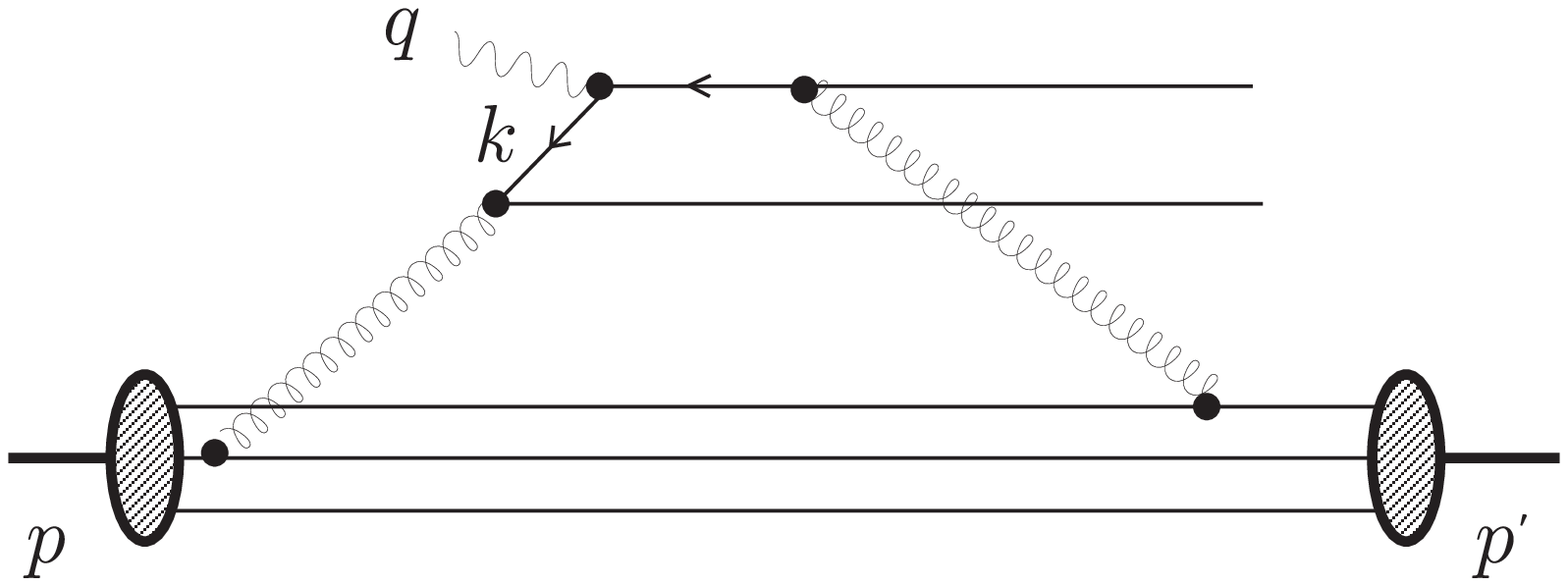 width 7 cm)}
\vspace{10pt}
\caption{Correspondence between the picture in the proton rest frame
(left) and that in a frame in which the proton has large momentum
(right).}
\label{F:correspondence}
\end{figure}

One of the primary motivations for examining diffractive DIS
experimentally is to use a short distance probe to study what
happens in diffractive scattering of a proton. The theoretical
language used to describe diffraction is Regge theory. One says
that the scattered proton exchanges a pomeron with the rest of the
system. We incorporate pomeron physics by writing the diffractive
parton distribution as
\begin{equation}
{ d f^{\rm diff}_{a/p}(\xi,x_\pom,t;\mu) \over dx_\pom\,dt}
=
{ |\beta(t)|^2 \over 8 \pi^2}\
x_\pom^{-2\,\alpha_\pom(t)}\
f_{a/\pom}(\xi/x_\pom,t;\mu).
\label{pomdist}
\end{equation}
Here $\beta(t)$ is the coupling of the pomeron to the proton and
$\alpha_\pom(t)$ is the pomeron trajectory function, $\alpha_\pom(0)
\approx 1$. The function $f_{a/\pom}(\xi/x_\pom,t;\mu)$ is called the
distribution of partons in the pomeron, although this name is perhaps
misleading in that it suggests that a pomeron is a kind of particle
that was emitted by the proton some time before the parton
distribution is probed. Eqs.~(\ref{difffact}) and (\ref{pomdist})
together constitute the Ingelman-Schlein model \cite{IngelmanSchlein}.

Eq.~(\ref{pomdist}) should be supplemented by additional terms
proportional to $x_\pom^{- 2\alpha_n(t)}$, with $\alpha_n <
\alpha_\pom$. One can neglect such subleading terms if $x_\pom$ is
small enough. The functions $f_{a/\pom}(\xi/x_\pom,t;\mu)$ are not
predicted by Regge theory (except in the small $\xi/x_\pom$ limit).
They provide experimental information on the nature of pomeron
exchange. Note that it is a prediction of the Regge theory that the
dependence on $x_\pom$ at fixed $\xi/x_\pom$ is $x_\pom^{-N}$ where
$N$ is independent of $\xi/x_\pom$.

We will see how Eqs.~(\ref{difffact}) and (\ref{pomdist}) compare to
experiment in the talk of Newman \cite{Newman}.

\section*{Theory of diffractive vector meson production}

\begin{figure}[htb]
\centerline{\DESepsf(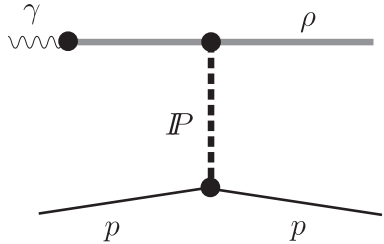 width 5 cm)}
\vspace{10pt}
\caption{Regge theoretic picture for photoproduction of vector
mesons.}
\label{F:vmd}
\end{figure}

Consider a collision between a proton and a real or nearly real
photon producing a diffractively scattered proton and vector meson
$V =\rho, \omega, J/\psi$:
\begin{equation}
\gamma + p \to V + p\,.
\end{equation}
There is no hard scale here, so perturbation theory does not apply.
However, one can describe the process using vector meson dominance
plus Regge theory, as depicted in Fig.~\ref{F:vmd}.

\begin{figure}[htb]
\centerline{\DESepsf(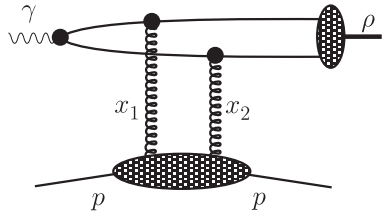 width 5 cm)}
\vspace{10pt}
\caption{Perturbative diagram for vector meson production with an
off shell photon.}
\label{F:rhoprod}
\end{figure}

If we now let the initial state photon be far off-shell, then we do
have a hard process, so that we can use a perturbative picture as
shown in Fig.~\ref{F:rhoprod}. The theory for this is discussed in
the talk of Nikolaev \cite{Nikolaev}. A nonperturbative function
$f(x_1,x_2)$ appears in Fig.~\ref{F:rhoprod}. This function is
closely related to the parton distribution function for finding a
gluon in a proton, $f_{g/p}(x)$, as discussed in the talks of
Radyushkin \cite{Radyushkin} and of Freund \cite{Freund}.

\section*{Hard diffraction in hadron-hadron collisions}

Consider the process $p+ \bar p \to p + jets + X$, where the jets
have a large transverse energy, so that we are looking at a hard
process. Similarly, we can consider diffractive $W$ boson production
and diffractive heavy quark production. As in the case of diffractive
DIS, we ask that the invariant momentum transfer $t$ between the
initial state proton and the final state proton obey $|t| \lesssim 1\
{\rm GeV}$ and that the proton's fractional longitudinal momentum
loss, $x_\pom$, is in the diffractive region $x_\pom \lesssim 0.05$.
Alternatively, we substitute a rapidity gap signal for the
observation of the diffracted proton.

\begin{figure}[htb]
\centerline{\DESepsf(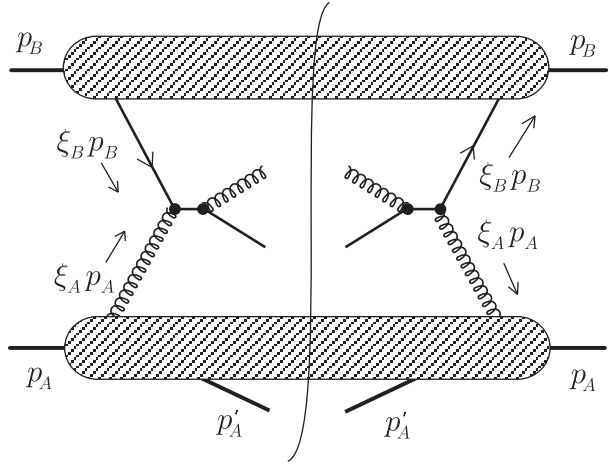 width 8 cm)}
\vspace{10pt}
\caption{Factored formula for diffractive jet production.}
\label{F:factoredjets}
\end{figure}

I show a simple picture for $p+ \bar p \to p + jets + X$ in
Fig.~\ref{F:factoredjets}. The corresponding formula is
\begin{equation}
{d \sigma^{\rm diff} \over dE_T\, dx_\pom\, dt}\sim
\sum_{ab}\int_0^{x_\pom}\!\!d\xi_A\ 
{d f^{\rm diff}_{a/A}(\xi_A;x_\pom,t;\mu) \over dx_\pom\,dt}
\int_0^1\!d\xi_B\ 
f_{b/B}(\xi_B;\mu)\
{d \hat\sigma_{ab} \over dE_T},
\label{factoredjets}
\end{equation}
where the diffractive parton distributions are the same ones that
appear in diffractive DIS. If we take the Regge form for these
functions, then Eq.~(\ref{factoredjets}) is the Ingelman-Schlein model
for diffractive jet production. The first indications in favor of
this model came from the UA8 experiment \cite{UA8}.

Should Eq.~(\ref{factoredjets}) work? There is good reason to believe
that the inclusive jet production cross section should take a
factored form \cite{CSSfactor}. However, with two hadrons in the
initial state, the argument requires a cancellation that occurs
precisely because we sum over all final states containing the required
jets. Here we do not sum over all such  final states: we demand that
the final state contain the diffractively scattered proton. Thus we
do not have an argument for factorization.  

\begin{figure}[htb]
\centerline{\DESepsf(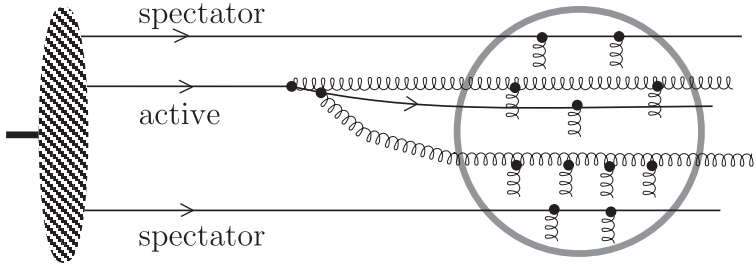 width 10 cm)}
\vspace{10pt}
\caption{Spectator parton effect leading to non-factorization.}
\label{F:survival}
\end{figure}

Furthermore, a simple argument shows that factorization should not be
expected. In Fig.~\ref{F:survival}, we look at the process from the
rest frame of the proton that is to be diffractively scattered. A
quark from hadron $B$, the ``active'' quark, initiates the jet
production long before hadron $B$ reaches hadron $A$. Since the quark
and the gluon that will form the jets have large transverse momentum,
they do not separate much before they pass through hadron $A$. A
second gluon, carrying a small fraction of the momentum of hadron $B$
and small transverse momentum is also produced. If we view the
process from a frame in which hadron $A$ has a large momentum, then
this gluon appears to be a constituent of hadron $A$. So far, we are
close to having a measurement of the gluon distribution in hadron
$A$. However, we have not accounted for the spectator quarks! (Also,
the jet system does not have the right color to correspond to a
measurement of the gluon distribution in hadron $A$. We could replace
it by a jet of the right color plus another spectator quark with the
color and transverse position of the active quark. Thus the color
issue is not different from the spectator quark issue.) In an
inclusive cross section, the spectator quarks would not matter.
However, for the diffractive cross section the interaction with the
spectator quarks can change the probability that hadron $A$ emerges
from the collision intact, having simply absorbed some momentum. It
seems likely that the extra interactions make it less likely for
hadron $A$ to survive. Thus we expect that the true cross section
will be given by the right hand side of Eq.~(\ref{factoredjets}) times
a survival probability $S$ that is less than 1. Unfortunately,
the survival probability is not calculable by perturbative methods.

The experimental results on diffractive hard scattering in $p \bar p$
collisions discussed in the working group on diffraction can be
summarized in the following table. In the first three rows, I show
the results for diffractive production of $W$ bosons, jets, and heavy
quarks, respectively. In each case, I show the rate for hard
diffractive scattering divided by the corresponding rate for
inclusive hard scattering. We see that these rates are generally a
fraction of a percent. In the right hand column, I show the
predictions based on Eq.~(\ref{factoredjets}). These results make use
of the diffractive parton distributions derived from HERA data by
Alvero {\it et al.}, as reported in talk of Whitmore \cite{Whitmore}.
That is, these results assume a suppression factor $S = 1$.
Evidently, the suppression factor is really something more like $S
\approx 0.1$.

\begin{table}[htb]
\caption{Results on diffractive hard scattering in $p \bar p$
collisions.}
\begin{tabular*}{5 in}{ll@{\hspace*{3em}}ll}
&&& \ \ \ \ If no \\
& \ \ \ \ CDF &\ \ \ \  D0 & suppression \\ \\
${{\rm Diffractive}\ W \over {\rm Inclusive}\ W}$ 
& (1.15 $\pm$ 0.55)\% 
&
&\ \ \ 9.4 \% \\ \\
${{\rm Diffractive\ Jets} \over {\rm Inclusive\ Jets}}$ 
& (0.75 $\pm$ 0.10)\% 
& 
&\ \ \  16 \% \\ 
&  
& (0.67 $\pm$ 0.05)\%
&\ \ \  10 \% \\ \\
${{\rm Diffractive}\ c,b \over {\rm Inclusive}\ c,b}$ 
& (0.18 $\pm$ 0.03)\% 
&
&\ \ \  ? \\ \\
${({\rm Diffractive})^2\ {\rm Jets} \over {\rm Inclusive\ Jets}}$ 
& $(2.7 \pm 0.7)\times 10^{-6}$ 
& $10^{-6}$
&\ \ \  ?\\ \\
\end{tabular*}
\label{T:diffhard}
\end{table}

In the fourth row of Table~\ref{T:diffhard}, I report the results
reported for the twice diffractive process $p + \bar p \to p + \bar p
+ jets + X$. Here both the proton and the antiproton are
diffractively scattered. The cross section is small, but still
observable.

One interesting connection between the $p \bar p$ data and the DIS
data emerged in the talk of Goulianos \cite{Goulianos}. If one assumes
Eq.~(\ref{factoredjets}), then one can get some information about
whether quarks or gluons predominate in the diffractive parton
distributions by comparing diffractive $W$ production (quark
dominated) to diffractive dijet production (gluon dominated). The
result is that gluons appear to predominate, in agreement with the
analysis based on HERA data. This result assumes that the survival
probability is $S = 1$, or, more generally, that the survival
probability is the same for both processes. Although this assumption
is open to question, the result is of interest because the gluons
participate directly in jet production, whereas the determination of
the diffractive gluon distribution in diffractive DIS is quite
indirect.

\section*{Jet-gap-jet events in hadron-hadron collisions}

\begin{figure}[htb]
\centerline{\DESepsf(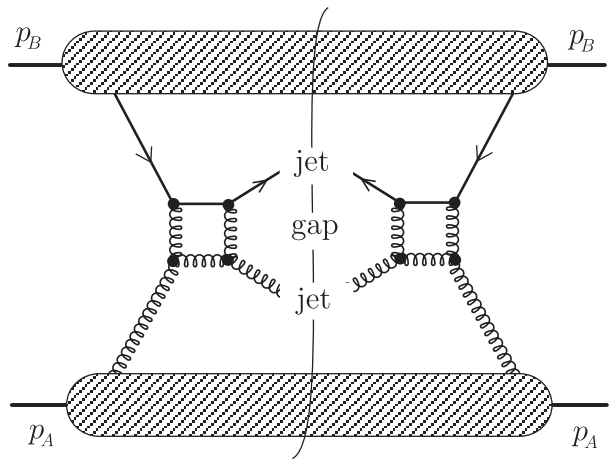 width 8 cm)}
\vspace{10pt}
\caption{Hard scattering via color singlet exchange leading to a
jet-gap-jet signature.}
\label{F:jetgapjet}
\end{figure}

We have discussed diffractive jet production in $p\bar p$ collisions,
in which one of the hadrons is diffractively scattered. If the
scattered hadron is not directly detected, then the signature for
this process is the existence of a rapidity gap in the detector in
the hadron fragmentation region. That is, there are no particles
found in the region of the detector at large rapidity, where the
fragments of the hadron would have been found if it had been broken
up. Consider now another kind of event, in which two jets are
produced and there are no particles found in the central rapidity
region lying between the two jets. We can call this a jet-gap-jet
event. Such an event can be produced by parton scattering via two
hard gluons, as illustrated in Fig.~\ref{F:jetgapjet}. The colors of
the two gluons should be such that the net color exchange is color
{\bf 1}. Then there is a right-moving, color singlet, system
consisting of hadron $A$ and one of the jets and there is a
left-moving color singlet system consisting of hadron $B$ and the
other jet. Soft color interactions would not be expected to produce
particles in the region between the left and right moving systems.
Thus there can be a gap containing no particles between the jets. Of
course, there can be a color exchange interaction between spectator
partons, so one expects a probability $S<1$ for the gap to
survive.

This process has been studied by CDF and D0, as reported in the talks
of Melese \cite{Melese} and Perkins \cite{Perkins} respectively. The
D0 study includes a comparison of the gap fraction at $\sqrt s  =
630\ {\rm GeV}$ to the gap fraction at $\sqrt s  =
1800\ {\rm GeV}$:
\begin{equation}
R =
{ [(\mbox{jet-gap-jet})/(\mbox{inclusive jet-jet})]_{630\ {\rm GeV}}
\over 
[(\mbox{jet-gap-jet})/(\mbox{inclusive jet-jet})]_{1800\ {\rm GeV}}}.
\end{equation}
The experimental result is $R = 2.6 \pm 0.6$. This can be compared to
the result predicted by the exchange of two hard gluons, as in
Fig.~\ref{F:jetgapjet}, $R= 0.8$. Better agreement is achieved in a
model in which the parton scattering occurs via one hard gluon
exchange and the color of the outgoing jets is rearranged by random
soft color interactions. This soft color rearrangement model predicts
$R=2.2$.

\bigskip
I thank the other convenors of the working group on diffraction,
P.~Newman, A.~Straiano, and P.~Melese, for their help in preparing this
talk.  I also thank Z.~Kunszt for discussions that have influenced
my thinking on the subjects covered here.
 


\begin{references}

\bibitem{Newman} P.~R.~Newman, these proceedings.

\bibitem{Nikolaev} N.~N.~ Nikolaev, these proceedings.

\bibitem{Wusthoff} M.~W\"usthoff, these proceedings.

\bibitem{Hebeker} A.~Hebeker, these proceedings.

\bibitem{Berera} A.~Berera, these proceedings.

\bibitem{IngelmanSchlein} G.~Ingelman and P.~Schlein, 
Phys.~Lett.~{\bf B152}, 256 (1985)..

\bibitem{BereraSoper} A.~Berera and D.~E.~Soper, 
Phys.~Rev.~D {\bf 50}, 4328 (1994);
Phys.~Rev.~D {\bf 53}, 6162 (1996).

\bibitem{Veneziano32} G.~Veneziano and L. Trentadue, 
Phys.~Lett.~{\bf B323}, 201 (1994).

\bibitem{Graudenz} D.~Graudenz, 
e-Print Achive:hep-ph/9701334.

\bibitem{Whitmore} J.~Whitmore, these proceedings.

\bibitem{Bartels} J.~Bartels, these proceedings.

\bibitem{Radyushkin} A.~Radyushkin, these proceedings.

\bibitem{Freund} A.~Freund, these proceedings.

\bibitem{UA8} R.~Bonino {\it et al.},
Phys.~Lett.~{\bf B211}, 239 (1988);
A.~Brandt {\it et al.},
Phys.~Lett.~{\bf B297}, 417 (1992).

\bibitem{CSSfactor} J.~C.~Collins, D.~E.~Soper, and G.~Sterman,
Nucl.~Phys.~{\bf B} 308, 833 (1986).

\bibitem{Goulianos} K.~Goulianos, these proceedings.

\bibitem{Melese} P.~Melese, these proceedings.

\bibitem{Perkins} J.~Perkins, these proceedings.

\end{references}
\end{document}